\documentclass[aps,prb,twocolumn,showpacs,superscriptaddress,amsmath,amssymb]{revtex4}

\usepackage{graphicx}        
\usepackage{color,colordvi}
\usepackage{epstopdf}
\usepackage{dcolumn}     
\usepackage{bm}              
\usepackage{psfrag}

\begin{document}
\title{Role of defect healing on the chirality of single-wall carbon nanotubes}
\author{M. Diarra}
\affiliation{Laboratoire d'\'Etude des Microstructures, ONERA-CNRS, BP 72, 92322 Ch\^atillon Cedex, France}
\affiliation{Centre Interdisciplinaire de Nanoscience de Marseille, CNRS and Aix Marseille University, Campus de Luminy, 13288  Marseille Cedex 09, France}
\author{H. Amara}
\affiliation{Laboratoire d'\'Etude des Microstructures, ONERA-CNRS, BP 72, 92322 Ch\^atillon cedex, France}
\author{C. Bichara}
\affiliation{Centre Interdisciplinaire de Nanoscience de Marseille, CNRS and Aix Marseille University, Campus de Luminy, 13288  Marseille Cedex 09, France}
\author{F. Ducastelle}
\affiliation{Laboratoire d'\'Etude des Microstructures, ONERA-CNRS, BP 72, 92322 Ch\^atillon cedex, France}
\date{\today}
\begin{abstract}

Although significant efforts have been directed towards a selective single wall carbon nanotube synthesis, the resulting diameter and chirality distributions are still too broad and their control remains a challenge. Progress in this direction requires an understanding of the mechanisms leading to the chiral selectivity reported by some authors. Here, we focus on one possible such mechanism and investigate the healing processes of defective tubes, at the atomic scale. We use tight-binding Monte Carlo simulations to perform a statistical analysis of the healing of a number of defective tubes. We study the role of temperature as a primary factor to overcome the energy barriers involved by healing, as well as the role of the metal catalyst. Using both electron diffraction patterns and local characterizations, we show that the healing proceeds first along the tube axis, before spreading laterally, and observe the competition between two or more chiralities. The resulting picture is that no chirality seems to be favored by the healing mechanisms, implying that the reported chiral preference should result from other sources.

\end{abstract} 


\maketitle

\section{Introduction} 

Owing to their exceptional electronic properties, single-walled carbon nanotubes (SWNTs) stand among potential materials for future electronic and optoelectronic technologies. Although significant efforts have been put on synthesis, the nanotubes diameter and chirality distributions are still too broad and their control remains quite challenging. Sorting techniques ~\cite{Hersam2008} such as ultracentrifugation in a gradient density ~\cite{Chen2007, Fleurier2009} or multicolumn gel chromatography ~\cite{Liu2011} are presently able to select nanotubes by diameters, chirality and also metallic or semiconducting character. However, such processes are complex and the tubes obtained are often defective and/or chemically spoiled. Such a structural control has also been reported to be achieved directly during the growth~\cite{Harutyunyan2009, Sundaram2011, Rao2012}. It has been shown that the SWNT chirality distribution obtained on supported CoMo catalysts can be influenced by experimental conditions where a majority of near-armchair (6,5) can be produced~\cite{Lolli06}. Recently, a link between the composition of Ni$_{x}$Fe$_{1-x}$ nanocatalysts and the resulting nanotube chirality has been observed~\cite{Chiang2009}. These experimental results are not always well understood because the detailed microscopic mechanisms involved in the synthesis of SWNTs are still lacking.

One does not know yet at which stage of the process, nucleation or growth, the diameter and chirality of the tube are fixed. One school of thought holds that the chirality is established during the nucleation stage. Static \textit{ab initio} calculations have been performed to compare various forms of possible nuclei on the surface of metal and establish which cap or capped tube is energetically preferred~\cite{Fan03, Reich06}. Reich \textit{et al.} suggested that the nucleation of the cap fixes the chirality of an individual tube as a change in chirality is unlikely during the growth phase~\cite{Reich06}. Yet, these calculations are performed on very crude atomic configurations of a cap sitting on a flat unrelaxed metal layer. \textit{In-situ} transmission electron microscopy (TEM) observations of the contact between the cap and the catalyst during the growth show a greater complexity suggesting that the chirality is established later during the growth~\cite{Hofmann2007, Yoshida2008, Hofmann2009, Pigos2011}. As an example, Hofmann \textit{et al.} have observed \textit{in-situ} the formation of a defected carbon nucleus from a metallic nanoparticle~\cite{Hofmann2007}, the development of a perfect tube being observed later. This implies that the defects observed at the initial stage have been annealed by some mechanisms, possibly involving the incorporation of additional C atoms into the network. Recent atomistic computer simulations by Neyts \textit{et al.}, using the so-called ReaxFF empirical potential, have confirmed this tendency~\cite{Neyts2011}. By performing hybrid reactive molecular dynamics/force-biased Monte Carlo simulations, the authors found that the chirality of the growing SWNT cap can change, indicating that the geometry of the tube is fixed later during the growth, in contradiction with the previous assumptions.

Computer simulation is a unique tool to analyze the onset of the SWNTs chirality at the atomic scale. Indeed, the atomic resolution is not yet obtained in the \textit{in-situ} TEM observation of the growth~\cite{Hofmann2007, Yoshida2008, Hofmann2009, Pigos2011}. Several models have been proposed to study the nucleation-growth mechanisms of SWNTs~\cite{Shibuta06, Shibuta03, Ding04, Ribas2009, Neyts10, Neyts2011, Shibuta2009, Page09, Amara08}. Nevertheless, whether the employed method is empirical~\cite{Shibuta06, Shibuta03, Ribas2009, Ding04, Neyts10, Neyts2011, Shibuta2009} or semi-empirical~\cite{Page09, Amara08}, all final configurations are plagued by a high concentration of atomic-scale defects, i.e. non hexagonal rings, adatoms or vacancies. A specific attention must then be paid to the defect healing mechanisms, to establish whether they can be held responsible for the reported chiral selectivity. We have recently shown that defects in graphene are highly sensitive to the temperature and discussed the key role played by metallic atoms in the reconstruction of a defected graphene sheet~\cite{Karoui10}.

In this paper, we investigate the healing processes of defective carbon SWNTs at atomic scale. We use a tight-binding model implemented in a Monte Carlo code to study the finite temperature evolution of defected tubes. Different lengths and diameters of nanotubes have been investigated at various temperatures. To understand the intrinsic character of the healing process we studied infinite tubes, using periodic boundary conditions, and no cap. The onset of the tube chirality is investigated and discussed through both electron diffraction patterns and local analyzes. The approach proposed here enables one to try and identify the healing mechanisms that produce perfect tube structures and those possibly favoring a definite chirality.

\section{Methodology}
\label{methodology}

As in our previous works, a tight-binding potential is used to model the interaction between metal (here nickel) and carbon atoms. To keep the model as simple and fast as possible, the local densities of electronic states are calculated using the moment's method where only the $s,p$ electrons of the carbon and $d$ electrons of the metal are taken into account. This model, both simple and accurate, is then implemented in a Monte Carlo (MC) code using either a canonical or a grand canonical (GC) algorithm with fixed volume, temperature, number of Ni atoms, and C chemical potential $\mu_{C}$. More details can be found in Ref.\ \cite{Amara09}. Using our carefully parameterized order $N$ tight-binding model, we have identified the temperature and carbon chemical potential conditions for the nucleation of a tube cap on nickel particles. We showed that  nucleation takes place after the outer Ni layer(s)  are saturated with carbon~\cite{Amara06, Amara08}. Recent technical improvements~\cite{Los11} of the algorithm of our tight-binding Monte Carlo code made it significantly faster. The exploration of longer time scales and more extensive investigations are now possible~\cite{Los10, Karoui10}.

\begin{figure}
\includegraphics[width=0.99\linewidth]{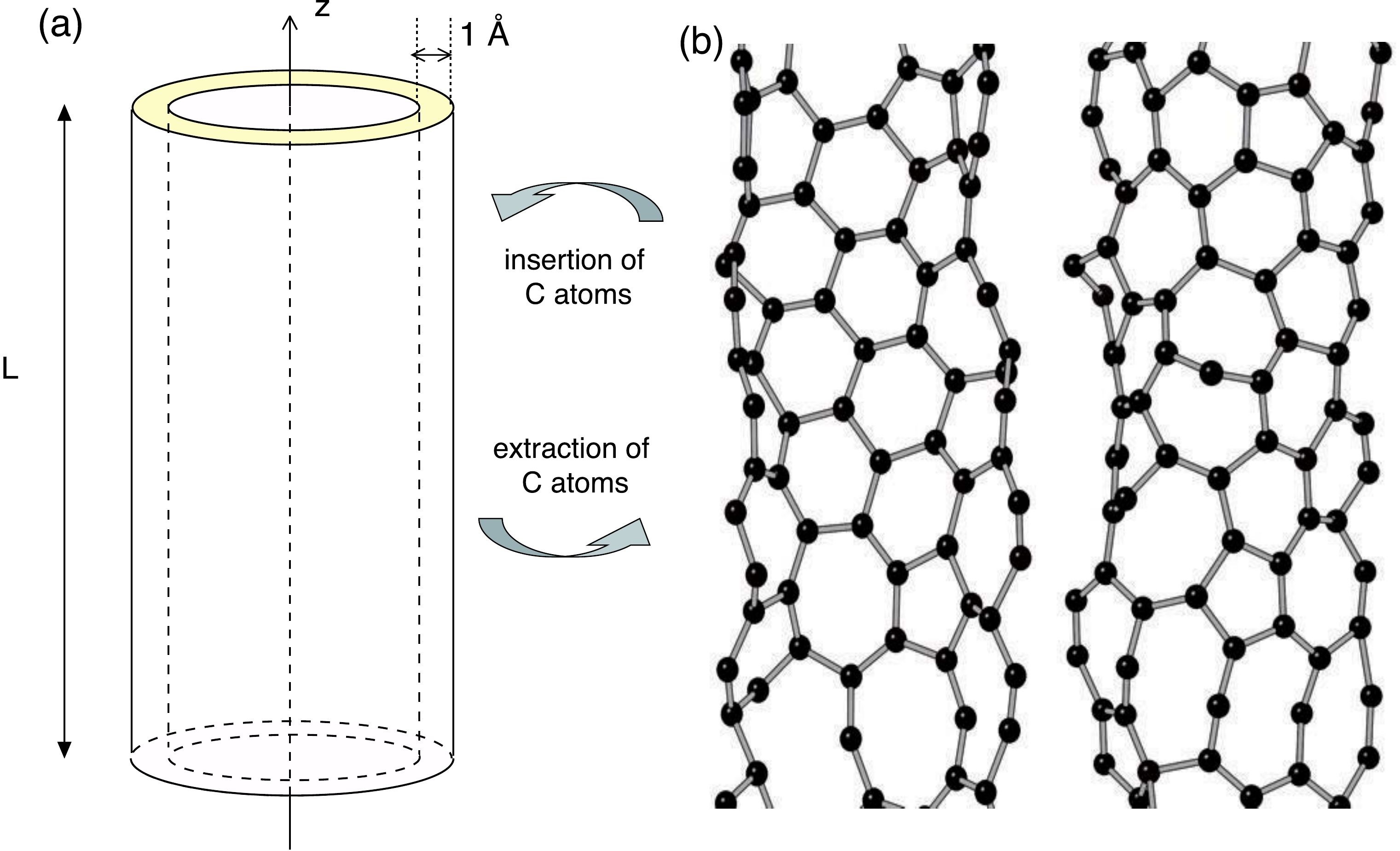}
\caption{(a) Schematic representation of the cylinder area where GCMC are performed to form defected tubes. (b) Two examples of defected tubes considered in the present study and obtained after performing GCMC calculations at 2500 K and a value of $\mu_{C}$ about -1.00 eV/atom.}
\label{Figure_1}
\end{figure}

We first generate a number of nanotubes with various realistic defects, by performing Grand Canonical Monte Carlo (GCMC) simulations where the active zone for inserting or removing carbon atoms is defined as a region of space around the surface of a cylinder (see Fig.~\ref{Figure_1}a). Moreover, periodic boundary conditions have been applied along the $z$ axis in order to avoid the closing of the tube. Indeed, due to the presence of dangling bonds, the spontaneous closing of the tube into a graphitic-like dome has been previously observed~\cite{Charlier1997}. The large number of defects, as shown in Fig.~\ref{Figure_1}b, results from the growth conditions: the carbon chemical potential chosen to obtain defected structures is unrealistically high (around -1.00 eV/at). Under such conditions, the growth is too fast and defects have a very high probability to be formed. In this way, we create a number of defected tubes with diameters and lengths ranging from 4.0 to 16 \AA~and 20 to 40 \AA~,respectively. The initial configurations (153 to 756 atoms) contain in average 60 \% of defects. The defects generally correspond to non hexagonal cycles and/or carbon atoms with a coordination number not equal to three. \\

\begin{figure}
\includegraphics[width=0.99\linewidth]{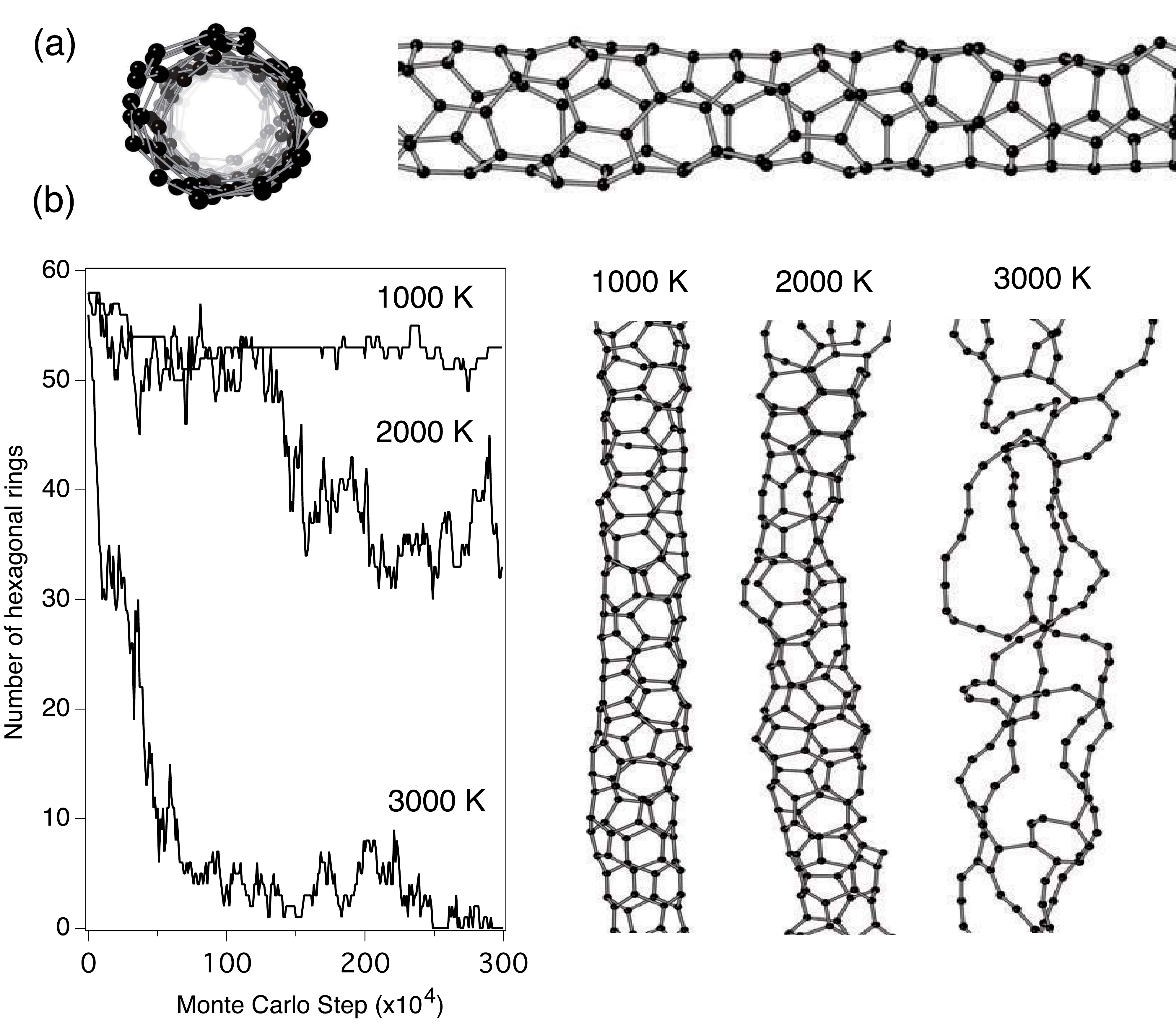}
\caption{(a) Defected tube with a diameter about 4 \AA.  (b) Variation of the number of hexagonal rings as function of Monte Carlo steps and equilibrium configurations at  1000, 2000 and 3000 K. }
\label{Figure_2}
\end{figure}

In a first step, we study the influence of the annealing temperature on the structure of a small tube with a diameter close to 4 \AA, as seen in Fig.~\ref{Figure_2}a. It corresponds to the narrowest attainable tube synthesized, using AlPO$_4$-5 as a template~\cite{Wang00, Roussel07}. The small tube, containing 180 C atoms, was subjected to temperatures ranging from 1000 to 3000 K (see Fig.~\ref{Figure_2}b). In this case, no healing was obtained. On the contrary, the number of defects increases with temperature until linear chains are formed for temperatures around 3000 K. This is not surprising because such small tubes are only marginally stable, due to the strong curvature of the wall, and have been observed only in the presence of a template.
\section{Healing mechanisms of defective nanotubes}

\subsection{Role of the temperature and chirality analysis}

\begin{figure}
\includegraphics[width=0.99\linewidth]{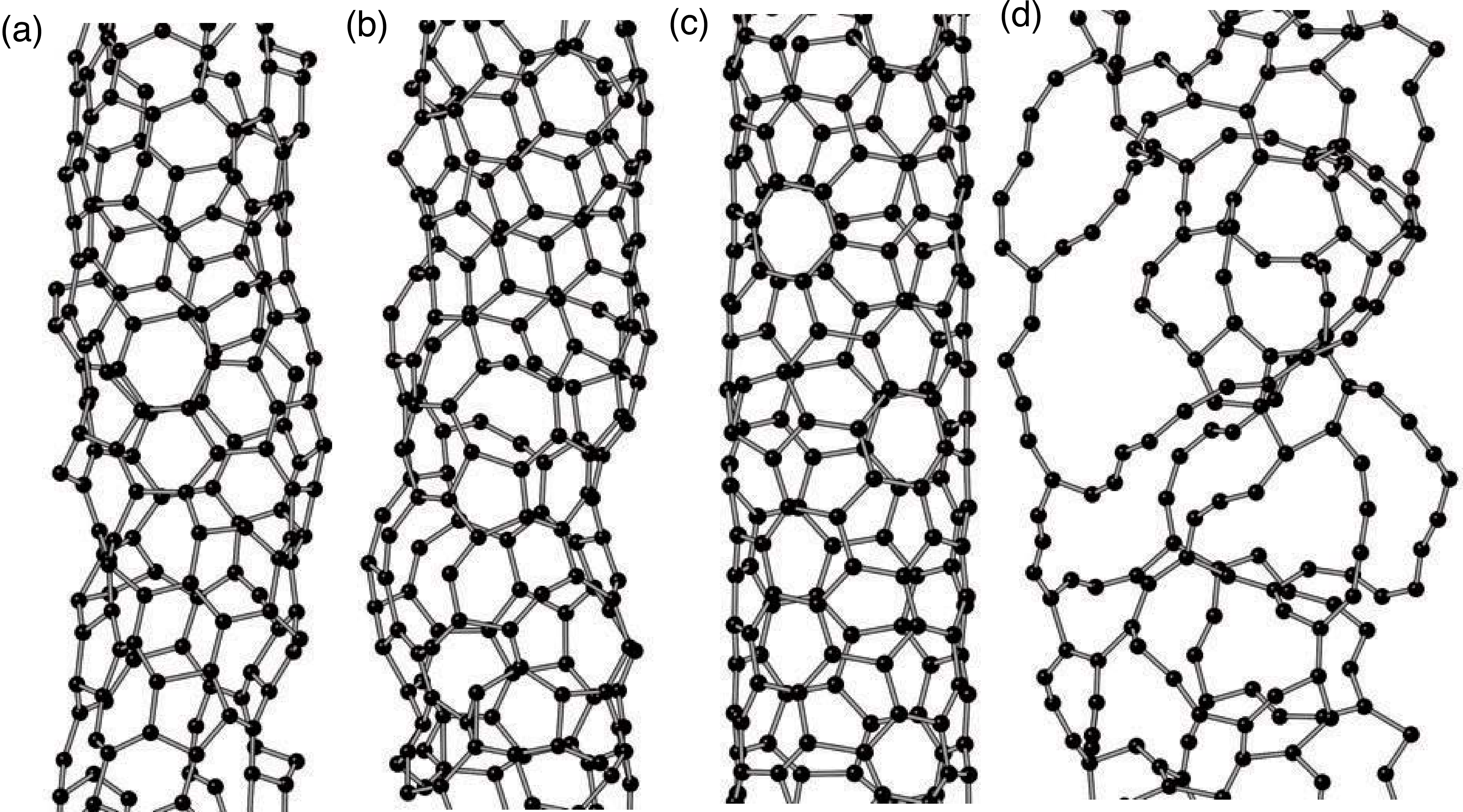}
\caption{Equilibrium configurations at 1500, 2000, 2500 and 3000 K.}
\label{Figure_3}
\end{figure}

\begin{figure}
\includegraphics[width=0.99\linewidth]{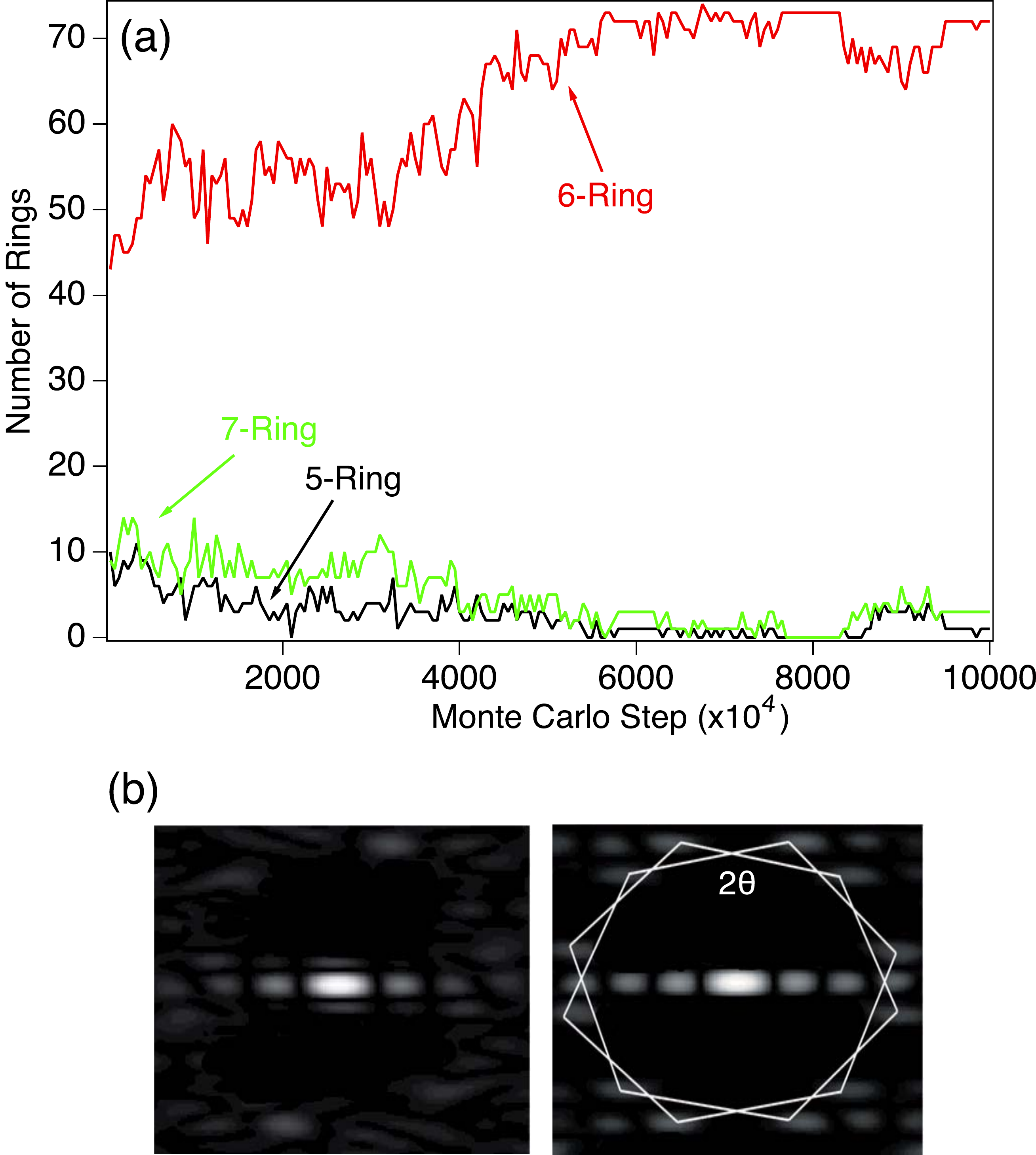}
\caption{(a) Variation of the number of polygons as function of Monte Carlo steps in a healed tube leading to a (6,4) at 2500 K. (b) Diffraction patterns of tubes before (left) and after (right) healing process. }
\label{Figure_4}
\end{figure}

We now focus on a larger defected nanotube whose diameter is between 6 and 8 \AA~with periodic boundary conditions along the $z$ axis (20 \AA). As in the previous case, this tube was subjected to temperatures ranging from 1000 to 3000 K in an attempt to heal the defects. The initial configuration contains 153 carbon atoms and 60~\% of the rings forming the tube walls are not hexagons. Figure~\ref{Figure_3} shows the final states of the MC runs at four temperatures (T$=$ 1500, 2000, 2500 and 3000 K). At 1500 and 2000 K, a significant amount of defects remains in the final configuration. Upon heating at 2500 K, the final configuration is an almost perfect tube where only 5 \% of the defects remain. As seen in Fig.~\ref{Figure_4}a, where the variation of the number of polygons during the MC simulation is presented, 70 hexagonal rings are present after the healing process. At higher temperature (T$=$ 3000 K), close to vaporization conditions, the tube is completely destroyed. The same conclusions hold for larger (diameters around 10 and 15 \AA) and longer (40 \AA) tubes studied. For systems without metal catalyst, 2500 K seems to be the optimal temperature, where the nanotube structure is relatively well healed.

To go further, we have characterized the chirality of the healed tubes. The detection of armchair or zigzag tubes is straightforward, but other cases require a more sophisticated method: we chose to calculate their electron diffraction patterns. This is a very powerful and reliable technique providing us with a direct determination of the chirality. SWNTs are commonly defined by the integer coordinates ($n$, $m$) of a so-called chiral vector $c_{n,m}=n\textbf{a}+m\textbf{b}$ in the 2D hexagonal lattice ($\textbf{a}$ and $\textbf{b}$ are unit vectors of the planar network). Given the diameter and the chiral angle $\theta$ of the tube, the chirality ($n$,$m$) can be determined. As an example, the comparison of the electron diffraction patterns produced by the defected and healed tubes studied previously is depicted in Fig.~\ref{Figure_4}a. All patterns have been obtained after performing simulated annealing to relax the structures. This step is necessary to decrease as much as possible the elongated shape of the spots and then facilitate the interpretation of the images. In the present case, we have found two hexagonal networks rotated from each other by twice the chiral angle $\theta \simeq$ 10.5. Since its diameter is around 6.68 \AA, we can assert that a (6,4) tube is observed after our healing process.
\begin{figure}
\includegraphics[width=0.99\linewidth]{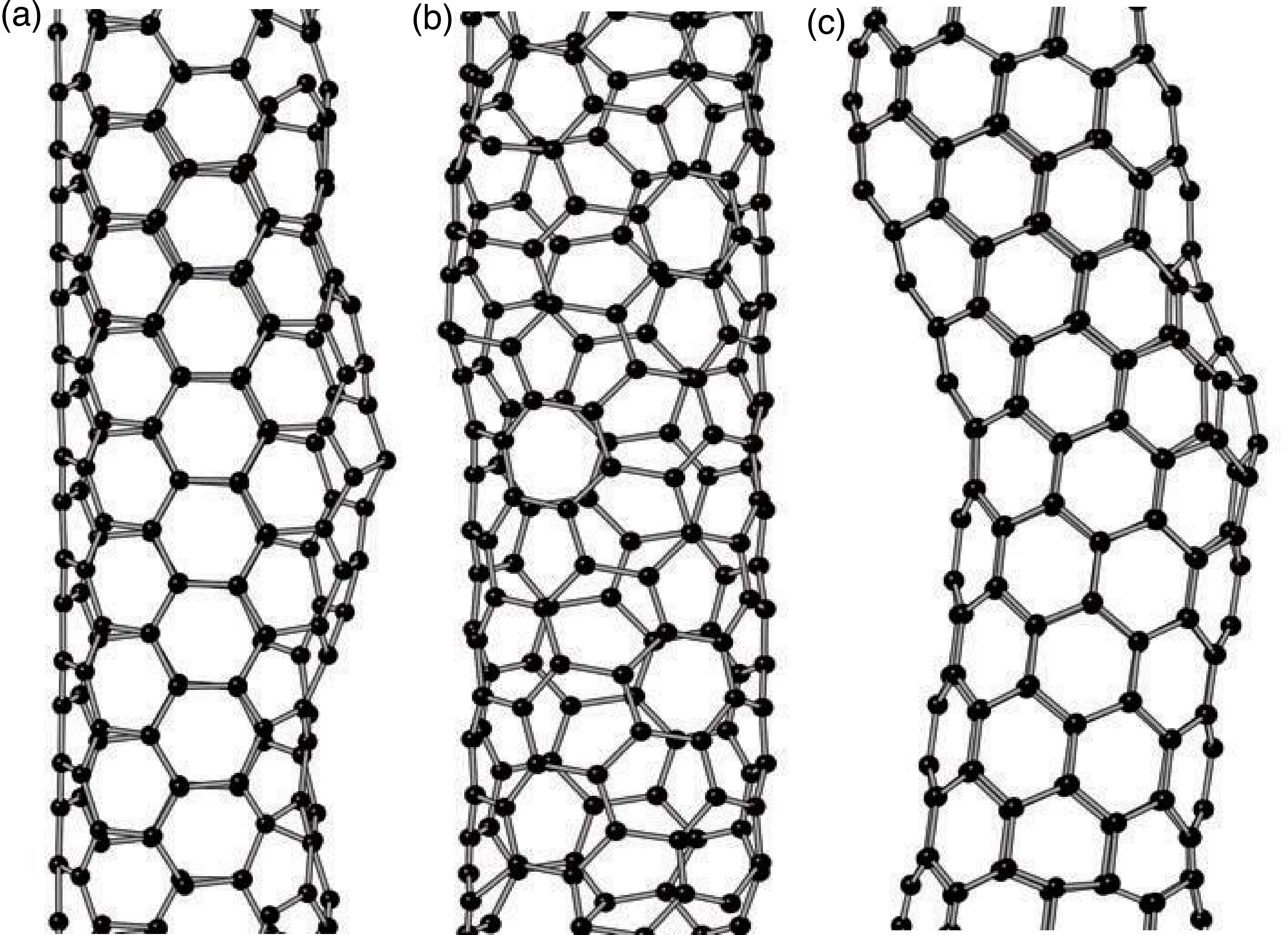}
\caption{Equilibrium configurations at 2500 K of healed tubes (a) (5,5), (b)(6,3), and (c) (5,5) and (9,0).}
\label{Figure_5}
\end{figure}

\begin{figure}
\includegraphics[width=0.99\linewidth]{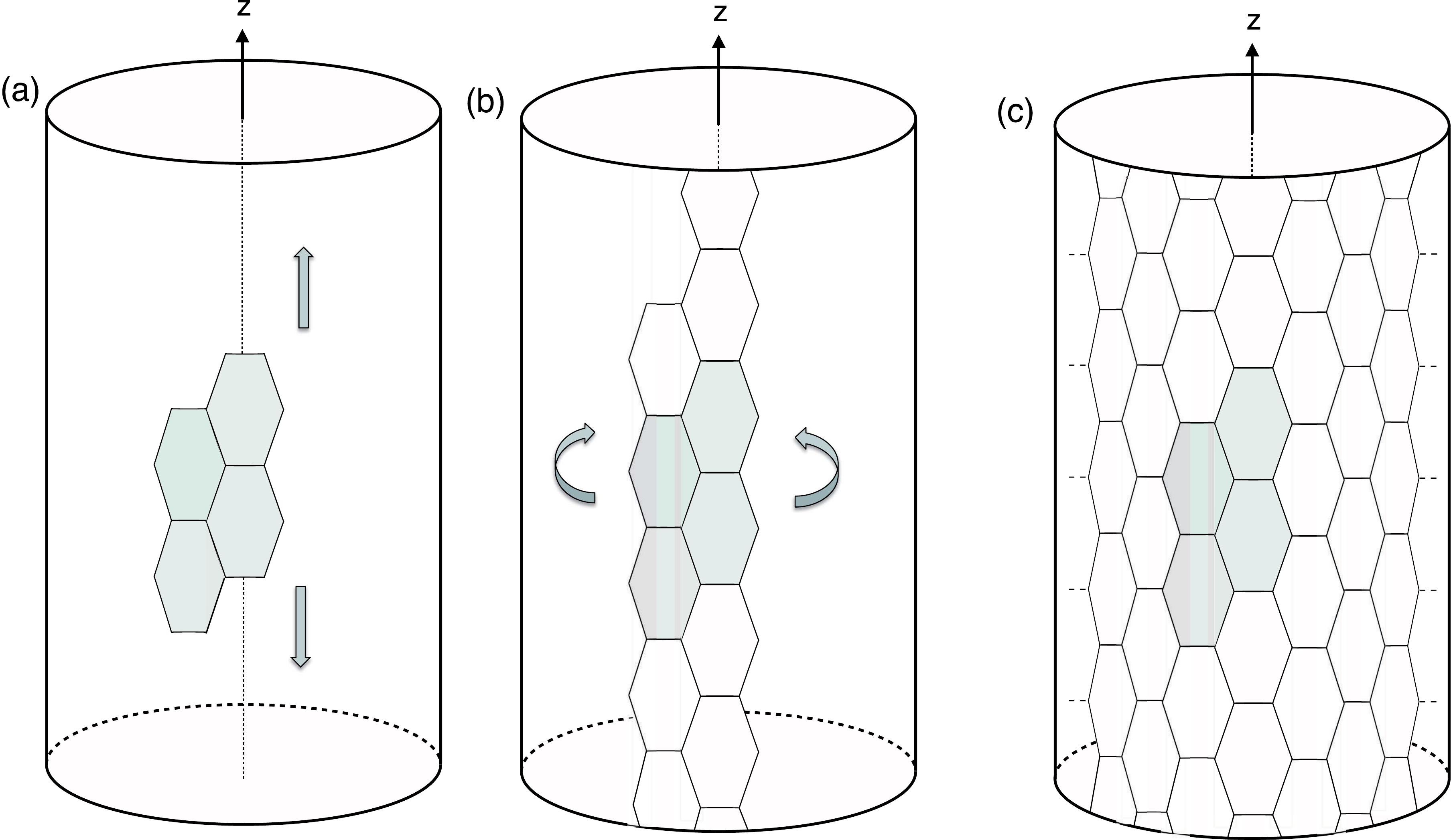}
\caption{Schematic representation of defect healing in the simulations.}
\label{Figure_6}
\end{figure}

\begin{figure}
\includegraphics[width=0.99\linewidth]{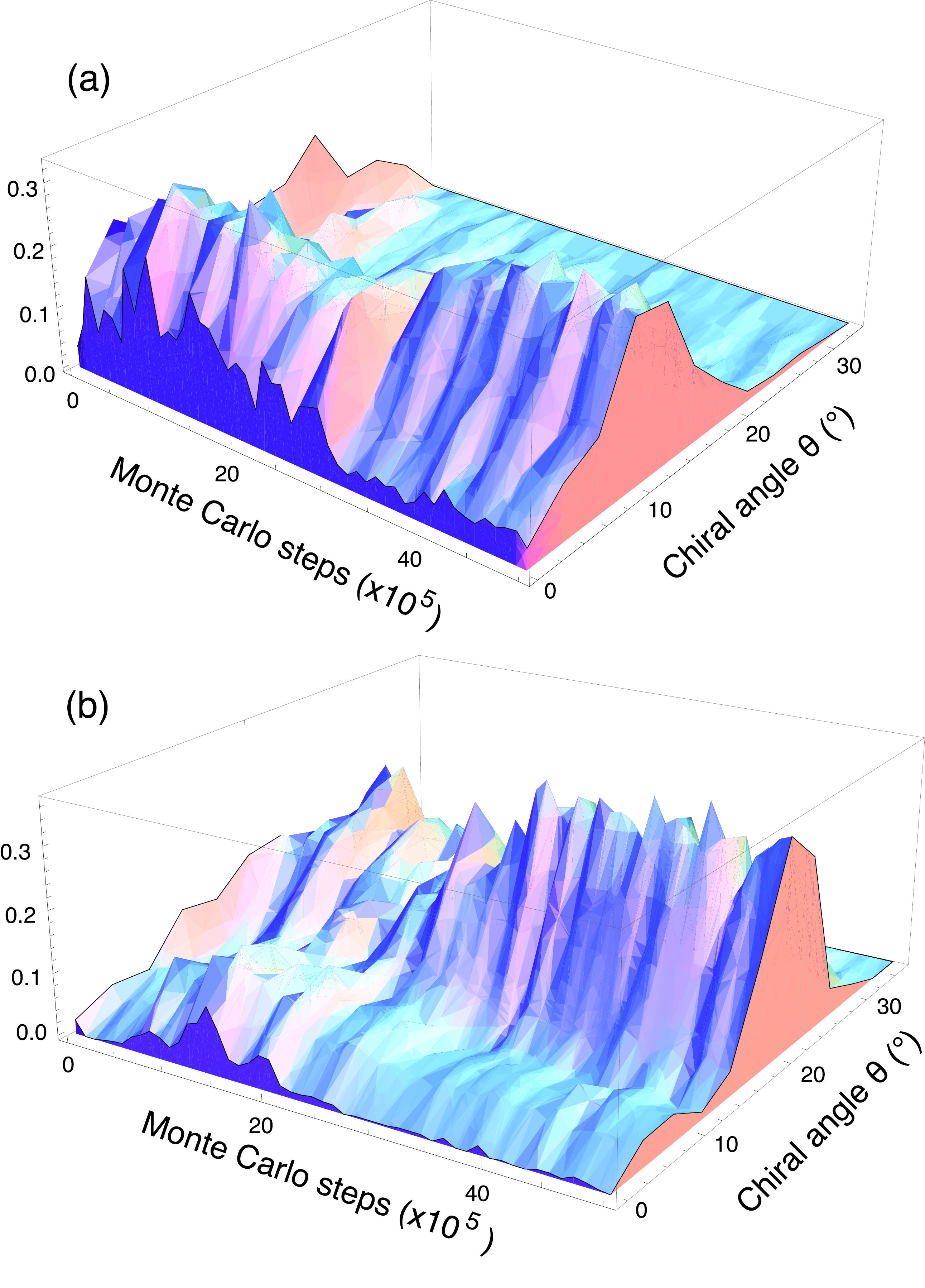}
\caption{Variation of the chirality as a function of the Monte Carlo steps at 2500 K. (a) One initial chirality kept all along the simulation.  (b) Competition between different seeds leading to one chirality at the end of the healing process.}
\label{Figure_7}
\end{figure}

To investigate how the tube's chirality builds up, we seek to identify the healing mechanism leading to almost perfect tubes. Our approach is to determine and follow the evolution of graphitic islands considered as seeds, with a local chirality based on the individual hexagons of the defected tubes. To distinguish a seed we have defined the following criterion. An area inside the defected tube is considered as a seed if at least four hexagons close together are observed. This threshold has been established in an empirical manner from our statistical observations during our numerous simulations. The observed healing process can be described as follows. Initially, the first seeds to be formed are destroyed by statistical fluctuations. When graphitic islands are sufficiently stable (Fig.~\ref{Figure_6}a), we have invariably observed their extension along the longitudinal axis of the tube leading to a ribbon of hexagons (see Fig.~\ref{Figure_6}b). To confirm this mechanism, we have considered defected structures with a belt of hexagons all around the tube and investigate its behavior at different temperatures. Instantaneously, most of the hexagonal rings forming the belt are destroyed and the process described previously can start again. Once the ribbons are formed (Fig.~\ref{Figure_6}b), the growth of the embryo proceeds transversely, leading to a  perfect tube as shown in Fig.~\ref{Figure_6}c. To determine when the tube structure is established, a local chirality has to be defined, instead of the global one used above. Such analysis has been performed thanks to an approach proposed by Kim \textit{et al.}~\cite{Kim2011}. Using this method, for ideal SWNTs at 0 K, the local chiral angles are exactly equivalent to the chiral angles of the tubes and the distribution histogram is single peaked, whereas at high temperatures it is broadened. In the case of defected SWNTs, the calculated local chiral angles show a multipeaked or broad distributions. If only one graphitic island with a defined chirality is present, the initial chirality is kept all along the process as shown in Fig.~\ref{Figure_7}a. In the case where several seeds with different chiralities are identified, there is a competition between the formation of structures with one (Fig.~\ref{Figure_7}b) or many chiralities. The latter case is depicted in Fig.~\ref{Figure_5}c where both (5,5) and (9,0) coexist until the end of the healing mechanism.

\subsection{Role of nickel}

\begin{figure*}
\includegraphics[width=0.99\linewidth]{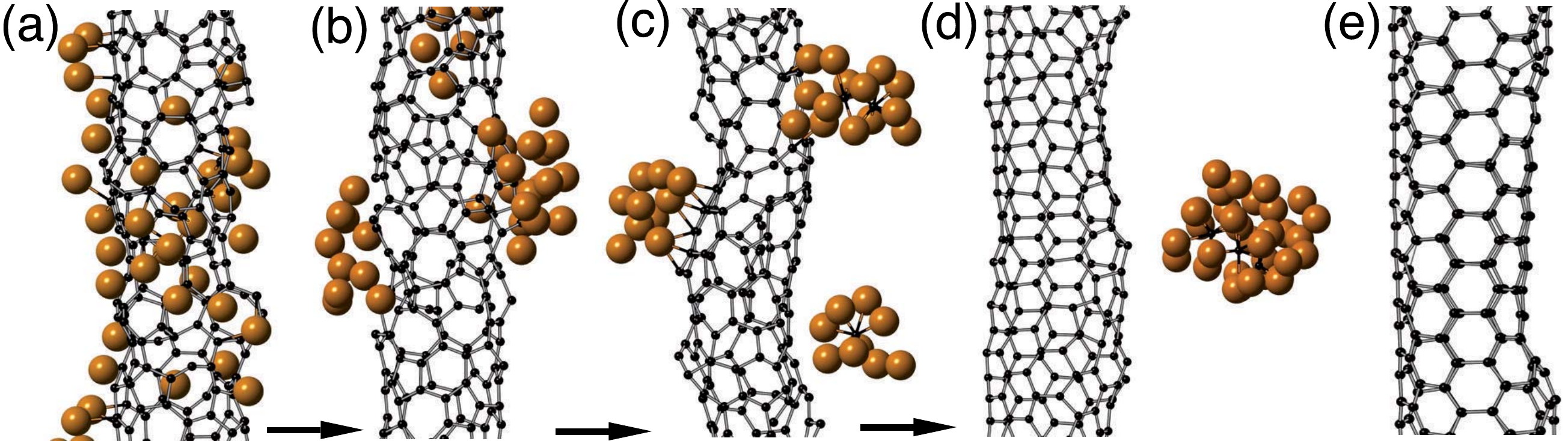}
\caption{(a)-(d) Healing process of defected tubes in the presence of Ni atoms. (e) Final configuration after relaxation (Ni atoms have been removed for the sake of clarity). }
\label{Figure_8}
\end{figure*}

We further investigate the correction and migration of defects in $sp^{2}$ structures by randomly adding Ni atoms on the walls of defective tube (Fig.~\ref{Figure_8}a). The aim of the present study is not to investigate the influence of a metallic nanoparticle in the growth process, which is very complex, but simply to understand the role played by metal atoms in the healing of defective carbon structures and also in the chiral selectivity. The system, containing 153 C and 35 Ni atoms, was subjected to temperatures ranging from 1000 to 2500 K. As in the case of the pure carbon structure, we notice that at 1000~K, the tube is far from being healed and a non-negligible fraction of defects remains. Once again, large rings are healed at 1500 and 2000 K, whereas pentagons and heptagons cancel out at higher temperatures. At 2500 K, the tube is completely healed as seen in Fig.~\ref{Figure_8}. This suggests that the healing of defects is thermally activated as observed for the isolated tube. By comparison with the case of tubes without catalyst, we deduce that Ni plays a crucial role in defect migration. In fact, during the first steps of the simulation, we have observed Ni atoms preferring to cluster together instead of distributing over the entire SWNT(Fig.~\ref{Figure_8}b). A similar behavior has been noticed in our previous simulations~\cite{Borjesson09} to explain the possibility of SWNT regrowth. As in the experiments where a vapor of metal atoms is deposited on the surface of SWNT seeds~\cite{Wang05}, we observed Ni atoms clustering at the lip of a rigid perfect tube to allow for regrowth. In the present instance, the nickel atoms tend to cluster and, when necessary, dissolve excess carbon atoms to get the right number of atoms to form a better healed tube. Upon heating at 2500 K, the final configuration is an almost perfect tube where only 2 \% of the defects remain. The key role of Ni atoms in catalytically healing defects carbon structures has already been observed with our TB model in the case of graphene~\cite{Karoui10} and by other groups in simulations on the growth of SWNT on Ni clusters~\cite{Page09, Neyts2011}.

\section{Favoured chirality}

To investigate a possible chiral selectivity resulting from the healing mechanisms, we performed a study of the healing of twenty defected tubes taken as starting configurations: ten isolated tubes and ten tubes with nickel atoms dispersed on. According to their diameter range ($\approx$ 6--8 \AA) and the periodic boundary conditions along the $z$ axis (20 \AA), the possible chiralities were: one armchair, (5,5), two zigzags, (8,0) and (9,0), and three chiral tubes, (6,3), (6,4) and (8,2). Within our model, the cohesive energy differences between these tubes is negligible ($\sim$ 1-2 meV/at), in agreement with \textit{ab initio} calculations~\cite{Sanchez1999, Kato2011}. According to the calculated energies, no perfect tube is energetically favored with respect to others. After performing MC simulations at 2500 K, all defected tubes are healed. For 80 \% of healed tubes, the exact determination of the chiralities through the calculation of diffraction patterns is straightforward since few defects are still present. After the healing process, all possible chiralities were randomly observed. Some examples of final equilibrium configurations are presented in Fig.~\ref{Figure_5}. The same conclusions hold for larger and longer tubes. Our statistical study reveals that no chiral selectivity linked to the healing mechanism is to be expected. This is in agreement with TEM investigations where the atomic structure of several tubes was determined precisely by combining a quantitative electron diffraction study and simulations based on kinematic theory~\cite{Henrard2000}.

\section{Conclusion}
To summarize, using tight-binding Monte Carlo simulations, we have investigated the healing processes of SWNTs to determine how and when the chirality is established. We have shown that high temperatures conditions are capable of healing very defective tubes. Using this approach, we observed that strongly defected tubes are first repaired along their longitudinal axis, before the remaining defects are taken care of along the tube circumference. Calculated electron diffraction patterns and local characterization of the chirality have been used to get a deeper insight into the onset of the SWNTs chirality during their synthesis. A statistical analysis demonstrates the formation of tubes with uniform diameter but random chirality. The intrinsic defect healing mechanisms studied here, that do not include the presence of a catalyst nanoparticle at the lip, seem then unlikely to favor a specific tube chirality. Other phenomena occurring at the tube / catalyst nanoparticle interface, and influencing the carbon atom incorporation energetics and kinetics, could play on the reported chiral selectivity and should be considered in future work.
\begin{acknowledgments}

The authors acknowledge the financial support of the GDRI-GNT (Graphene and Nanotubes) and the ANR (ANR-09-NANO-028 : ``SOS Nanotubes'').

\end{acknowledgments}

\begin{thebibliography}{0}
\expandafter\ifx\csname natexlab\endcsname\relax\def\natexlab#1{#1}\fi
\expandafter\ifx\csname bibnamefont\endcsname\relax
  \def\bibnamefont#1{#1}\fi
\expandafter\ifx\csname bibfnamefont\endcsname\relax
  \def\bibfnamefont#1{#1}\fi
\expandafter\ifx\csname citenamefont\endcsname\relax
  \def\citenamefont#1{#1}\fi
\expandafter\ifx\csname url\endcsname\relax
  \def\url#1{\texttt{#1}}\fi
\expandafter\ifx\csname urlprefix\endcsname\relax\def\urlprefix{URL }\fi
\providecommand{\bibinfo}[2]{#2}
\providecommand{\eprint}[2][]{\url{#2}}

\end{thebibliography}


\begin{thebibliography}{99}
%
\bibitem{Hersam2008}
M. C. Hersam, Nat. Nano.  \textbf{3}, 387 (2008)
%
\bibitem{Chen2007}
F. Chen, B. Wang, B, Y. Chen, and L.-J Li, Nano. Lett. \textbf{7},  3013 (2007)
%
\bibitem{Fleurier2009}
R. Fleurier, J.-S Lauret, U. Lopez, and A. Loiseau, Adv. Func. Mat. \textbf{19}, 2219 (2009)
%
\bibitem{Liu2011} 
H. Liu, D. Nishide, T. Tanaka, and H. Kataura, Nat. Comm. \textbf{2}, 309 (2011)
%
\bibitem{Harutyunyan2009}
A. R. Harutyunyan, G. Chen, T. M. Paronyan, E. M. Pigos, O. A. Kuznetsov, K. Hewaparakrama, S. M. Kin, D. Zakharov, E.A. Stach, and G. U. Sumanasekera, Science \textbf{326}, 116 (2009).
%
\bibitem{Sundaram2011}
R. M. Sundaram, K. K. K Koziol, and A. H. Windle, Adv. Mat. \textbf{23}, 5064 (2011).
%
\bibitem{Rao2012}
R. Rao, D. Liptak, T. Cherukuri, and B. I. Yakobson, B. Maruyama, Nat. Mater. \textbf{11}, 213 (2012).
%
\bibitem{Lolli06}
G. Lolli, L. Zhang, L. Balzano, N. Sakulchaicharoen, Y. Tan, and D. E. Resasco, J. Phys. Chem. B \textbf{110}, 2108 (2006).
%
\bibitem{Chiang2009}
W. -H. Chiang and R. Mohan Sankaran, Nat. Mater. \text{8}, 882 (2009).
%
\bibitem{Fan03}
X. Fan, R. Buczko, A. A. Puretzky, D. B. Geohegan, J. Y. Howe, S. T. Pantelides, and S. J. Pennycook, Phys. Rev. Lett. \textbf{90}, 145501 (2003).
%
\bibitem{Reich06}
S. Reich, L. Li, and J. Robertson, Chem. Phys. Lett. \textbf{421}, 469 (2006).
%
\bibitem{Hofmann2007}
S. Hofmann, R. Sharma, C. Ducati, G. Du, C. Mattevi, C. Cepek, M. Cantoro, S. Pisana, A. Parvez, F. Cervantes-Sodi, A. C. Ferrari, R. Dunin-Borkowski, S. Lizzit, L. Petaccia, A. Goldoni, and J. Robertson, Nano. Lett. \textbf{7}, 602 (2007).
%
\bibitem{Yoshida2008}
H. Yoshida, S. Takeda, T. Uchiyama, H. Kohno, and Y. Homma, Nano. Lett. \textbf{8}, 2082 (2008).
%
\bibitem{Hofmann2009}
S. Hofmann, R..Blume, C. T. Wirth, M. Cantoro, R. Sharma, C. Ducati, M. H\"avecker, S. Zafeiratos, P. Schnoerch, A. Oestereich, D. Teschner, M. Albrecht, A. Knop-Gericke, R. Schl\"ogl, and J. Robertson, J. Phys. Chem. C \textbf{113}, 1648 (2009). 
%
\bibitem{Pigos2011}
E. Pigos, E. S. Penev, M. A. Ribas, R. Sharma, B. I.Yakobson, and A. R. Harutyunyan, ACS Nano \textbf{5}, 10096 (2011).
%
\bibitem{Neyts2011}
E. C.Neyts, A. C. T. van Duin, and A. Bogaerts, J. Am. Chem. Soc. \textbf{133}, 17225 (2011).
%
\bibitem{Shibuta2009}
Y. Shibuta and J.A. Elliott, Chem. Phys. Lett. \textbf{472}, 200 (2009). 
%
\bibitem{Shibuta06}
Y. Shibuta and J.A. Elliott, Chem. Phys. Lett. \textbf{427}, 365 (2006).
%
\bibitem{Shibuta03}
Y. Shibuta and S. Maruyama, Chem. Phys. Lett. \textbf{382}, 381 (2003).
%
\bibitem{Ding04}
F. Ding, A. Ros\'en, and K. Bolton, J. Chem. Phys. \textbf{121}, 2775(2004).
%
\bibitem{Ribas2009}
M. A. Ribas, F. Ding, P.B. Balbuena, and B. I. Yakobson, J. Chem. Phys. \textbf{131}, 224501(2009).
%
\bibitem{Neyts10}
E. C. Neyts, Y. Shibuta, A. C. T. van Duin, and A. Bogaerts, ACS Nano \textbf{4}, 6665 (2010).
%
\bibitem{Page09}
A. J. Page, Y. Ohta, Y. Okamoto, S. Irle, and K. Morokuma, J. Chem. Phys. C \textbf{113}, 20198 (2009).
%
\bibitem{Amara08}
H. Amara, C. Bichara, and F. Ducastelle, Phys. Rev. Lett. \textbf{100}, 056105 (2008).
%
\bibitem{Karoui10}
S. Karoui, H. Amara, C. Bichara, and F. Ducastelle, ACS Nano \textbf{4}, 6114 (2010).
%
%
\bibitem{Amara09}
H. Amara, J. -M. Roussel, C. Bichara, J. -P. Gaspard, and F. Ducastelle, Phys. Rev. B , \textbf{79}, 014109 (2009).
%
\bibitem{Amara06}
H. Amara, C. Bichara, and F. Ducastelle, Phys. Rev. B, \textbf{73}, 113404 (2006).
%
\bibitem{Los11}
J. H. Los, C. Bichara, and R. J. M. Pellenq, Phys. Rev. B \textbf{84}, 085455 (2011).
%
\bibitem{Los10}
J. H. Los, and R. J. M.  Pellenq, Phys. Rev. B \textbf{81}, 064112 (2010).
\bibitem{Charlier1997}
J. -C. Charlier, A. De Vita, X. Blase, and R. Car, Science \textbf{31}, 647 (1997).
%
\bibitem{Wang00}
N. Wang, Z. K. Tang, G. D. Li, and J. S. Chen, Nature \textbf{408}, 50 (2000).
%
\bibitem{Roussel07}
T. Roussel,  R. J.-M. Pellenq and C. Bichara, Phys. Rev. B \textbf{76}, 235418 (2007).
%
\bibitem{Kim2011}
J. Kim, S. Irle, and K. Morokuma, Phys. Rev. Lett. \textbf{107}, 175505 (2011).
%
\bibitem{Borjesson09}
A. B\"orjesson, W. Zhu, H. Amara, C. Bichara, and K. Bolton, Nano. Lett. \textbf{9}, 1117 (2009).
%
\bibitem{Wang05}
Y. Wang, M. J. Kim, H. Shan, C. Kittrell, H. Fan, L. M. Ericson, W.-F. Hwang, S. Arepalli, R. H. Hauge, and R. E. Smalley, Nano. Lett. \textbf{5}, 997 (2005).
%
\bibitem{Sanchez1999}
D. S\'anchez-Portal, E. Artacho, J. M. Soler, A. Rubio, and P. Ordej\'on, Phys. Rev. B \textbf{59}, 12678 (1999).
%
\bibitem{Kato2011}
K. Kato and S. Saito, Physica E \textbf{43}, 669 (2011).
%
\bibitem{Henrard2000}
L. Henrard, A. Loiseau, C. Journet, and P. Bernier, Eur. Phys. J. B \textbf{13}, 661 (2000).
%
\end{thebibliography}
\end{document}